\documentclass[aps,pre,reprint,10pt,superscriptaddress,showpacs]{revtex4-1}
\usepackage{amsfonts} 
\usepackage{amsmath}
\usepackage{amssymb}
\usepackage{graphicx}%\[  \]
\usepackage{subfigure}
\usepackage{color}
\usepackage{caption}
\begin{document}
\title{Synchronization in networks of coupled hyperchaotic CO$_2$ lasers}
\author{Animesh Roy}
\email{aroyiitd@gmail.com}
\affiliation{Department of Mathematics, Siksha Bhavana, Visva-Bharati (A Central University), Santiniketan-731 235, West Bengal, India}
\author{A. P. Misra}
\email{apmisra@visva-bharati.ac.in}
\affiliation{Department of Mathematics, Siksha Bhavana, Visva-Bharati (A Central University), Santiniketan-731 235, West Bengal, India}
\author{Santo Banerjee}
\email{santoban@gmail.com}
\affiliation{Institute for Mathematical Research, Universiti Putra Malaysia, Selangor, Malaysia}

%\author{A. P. Misra}
%\address{Department of Mathematics, Siksha Bhavana, Visva-Bharati University, Santiniketan-731 235, West Bengal, India}
%\eadd{apmisra@visva-bharati.ac.in; apmisra@gmail.com}
%\vspace{10pt}
%\begin{indented}
%\item[] 9 Jan 2020  
%\end{indented}
%%%%%%%%%%%%%%%%%%%%%%%%%
%\keywords{} $CO_2$ laser, Hyperchaos, Synchronization,Master stability function.
 
\begin{abstract}
We propose a non-autonomous dynamical system for an optically modulated CO$_2$ laser                                                                                                                                                                                                                                                                                                                                                                                                                                                                                                                                                                                                                                                                                                                                                                                                                                                                                                                                                                                                                                                                                                                                                                                                                                                                                                                                                                                                                                                                                                                                                                             and show that  it exhibits hyperchaos in presence of electro-optic feedback beams.   The  system is then used to study the synchronization in networks of mutually coupled hyperchaotic CO$_2$ lasers. By the method of  master stability function (MSF)  it is shown  that the   stable synchronous state  can be reached  for both the  ring of diffusively coupled (RDC) and star-coupled (SC) networks  of at most  $24$ nodes or oscillators.   However, in the former  networks, high-coupling strengths   $(\sim10)$ are required for synchronization compared to the latter ones $(\sim1)$.  A   numerical simulation of the coupled $24$ hyperchaotic CO$_2$ lasers is also performed to show that    the  corresponding synchronization error   $\lesssim10^{-6}$. Furthermore, the chimera states of the networks are found to coexist in  some intervals of time and the coupling strengths where the networks are not synchronized, implying that the synchronization occurs only in some specific ranges of values of the coupling strengths.  
\end{abstract}
\maketitle
\section{Introduction} \label{sec-intro}
Modulated CO$_2$ lasers have been known to be one of the  simplest, most useful  and efficient   laser systems  for applications in science and engineering, as well as  for various theoretical investigations \cite{bonatto2005}. After the pioneering work of Arecchi \textit{et al.}  \cite{arecchi1982}, who dealt with the measurement of subharmonic bifurcations, multistability, and chaotic behaviors in a Q-switched CO$_2$ laser,  CO$_2$ lasers have been  fruitfully explored in many directions, e.g., in communication systems \cite{olson1995},    stochastic bifurcations in modulated  CO$_2$ lasers \cite{billings2004},      neural networks \cite{liu2014}, fabrication of helical long-period gratings   in a polarization-maintaining fiber using CO$_2$ lasers \cite{jiang2018}.
\par
Over the last $30$ years, the CO$_2$ laser was extensively studied theoretically, numerically, and experimentally focusing  mainly on its   dynamical properties in phase space formalism \cite{gilmore1998,pisarchik2001}.  CO$_2$ lasers   have been experimentally shown to exhibit chaos due to delayed feedback, coupling with other  CO$_2$ lasers etc.  Such chaotic features have been shown to be controlled by using a modified proportional feedback technique \cite{perez1994,ciofini1999} and a negative feedback of subharmonic components of laser  intensity signal \cite{meucci1997}.  Furthermore, the possibility of the existence of chaos and its control to exhibit periodic orbits or steady states in nonlinear dynamical systems by means of small-amplitude perturbations has opened up new aspects in nonlinear dynamics both from a theoretical point of view \cite{ott1990} and when  it comes to applications \cite{hunt1991}.  
\par 
 The dynamics of coupled nonlinear systems has gained much interests in recent times because of their   spatiotemporal behaviors and related synchronization phenomena in theoretical physics and other fields of science \cite{pecora1990,kocarev1995}.  Furthermore, networks of dynamical systems are common   in many branches of science and engineering, and the social sciences \cite{parlitz1996,militello2018,hu2012,zheng2011}. 
In networks of coupled dynamical systems or oscillators, the strongest form of their cooperative dynamics is the synchronization, and  some interesting features can occur when all the subsystems behave in the same fashion. Such behavior of a network, models various continuous dynamical systems that have uniform movement, as well as electronic circuits,  neurons   and coupled lasers that synchronize.  Typically,  two stable systems are said to be synchronized, i.e., they do the same thing at the same time, when    their time evolution is periodic with the same period and maybe the same phase. However, this scenario changes   when the systems are chaotic \cite{li2019a}, and especially hyperchaotic \cite{li2018}.   In this context, a number of works has been proposed   in chaotic dynamical systems \cite{he2018,he2016} which have considered the synchronization in large networks of coupled systems with different coupling configurations. Furthermore, the conditions for the complete synchronization, i.e., under what conditions the stability of the synchronous state   occurs, especially with  the coupling strength and coupling configurations of the network, have also been studied  in various networks of periodic \cite{somers1995} and chaotic dynamical systems  (see, e.g., Refs. \cite{heagy1994a,heagy1994b,barahona2002,tang2019,karimi2019}). 
Much attention has also been paid to  inspect the correspondence between synchronization of oscillators forming  networks and the network topology. The latter plays a significant role in network synchronization as densely coupled networks   synchronize easily compared to the sparse networks \cite{belykh2004}.  
%In this regard,    various stability methods with algebraic, statistical, and graph approaches have been developed to understand what factors are responsible for the onset of synchronization in a given network of oscillators.  
\par 
Typically,   the synchronous solution in networks of continuous time oscillators   becomes stable when the coupling strength between the oscillators exceeds a critical value. This critical value depends on the individual oscillator dynamics and on the network topology.  However, the main concern is   to find the bounds for the coupling strengths for which the stability of synchronization is assured. To resolve this issue, a master stability function (MSF) has been proposed by Pecora \textit{et al.} \cite{pecora1998} which can be used to one's choice of stability requirement. The MSF relies on the calculation of the maximum Lyapunov exponent   for the least stable transverse modes of the synchronous manifold along with the eigenvalues of the connectivity matrices.  
\par 
The purpose of the present work is to propose a dynamical system for  CO$_2$ lasers which exhibit  hyperchaos in presence of optically modulated  feedback beams, one in the form of a small-amplitude time-dependent perturbation and the other a negative feedback of subharmonic components of the signal intensity driven by a voltage $V$.  Next,  we study the synchronization of a network  of  a large number of coupled hyperchaotic CO$_2$ lasers by the method of MSF as proposed by Pecora \textit{et al.} \cite{pecora1998}. We show that the   synchronous state can be stable for a longer time for both the  ring of   diffusively (nearest-neighbor) coupled (RDC) and star-coupled (SC) networks of at most $24$ nodes. However, in the former networks the coupling strengths need to be high $(\sim10)$ compared to the latter ones $(\sim1)$.  A numerical simulation of the coupled CO$_2$ lasers is also performed to show that  the  synchronization error  for   $24$  nodes     is   $\lesssim10^{-6}$. 
\section{The model and its dynamical properties} 
We propose a theoretical model for CO$_2$ lasers  that includes the combined effects of the injected feedback beam in the form of a small-amplitude time dependent perturbation and the negative feedback of laser intensity driven by a voltage $V$, and thus   modifies  the models   in Refs.    \cite{perez1994,meucci1997} . The equations governing the dynamics of CO$_2$ lasers are \cite{perez1994,meucci1997}
\begin{equation}
\begin{split}
\frac{dI}{dt}&=-Ik(V)-k_0 I \epsilon \cos{\Omega t} +\alpha IN,\\
\frac{dN}{dt}&=-\gamma(N-N_0)-2\alpha IN,\\
\frac{dV}{dt}&=-\beta\left(V-B+\frac{RI}{1+\eta I}\right),  \label{eq-basic-dim}
\end{split}
\end{equation}
%\frac{dZ}{dt}&=\Omega,\\
where $I$ is the dimensionless laser intensity,  $N$ is the population inversion, $\alpha$ is the coupling coefficient for $I$ and $N$, $\gamma$ is the population inversion decay constant, and  $N_0$ is the pumping power. The intensity decay rate   $k(V)$   of the cavity, which  depends on the voltage $V$ and is to obtained through a feedback loop, is given by
\begin{equation}
k(V)=k_0\left[1+k_1\sin^2\left(\frac{\pi(V-V_0)}{V_\lambda}\right)\right].\label{eq-k-V}
\end{equation}
Here, $k_0$ and $k_1$ are constants which depend on the cavity length and the total transmission for a single pass, and $V_0$ and $V_\lambda$ are constants associated with an offset and half-wave voltage of the modulator respectively. Also, $B$ is the   control parameter associated with the bias voltage,  $\beta$ is the damping rate, and $R$ is the total gain of the feedback loop such that $\eta R$ accounts for the nonlinearity of the detection aparatus. For more details of the discussion on different parameters readers are referred to Refs. \cite{perez1994,meucci1997}.
 %%%%%%%%%%%%%%%%%%
 \begin{figure*}[h!]
  \begin{center}  
        \includegraphics[width=6in,height=2.5in]{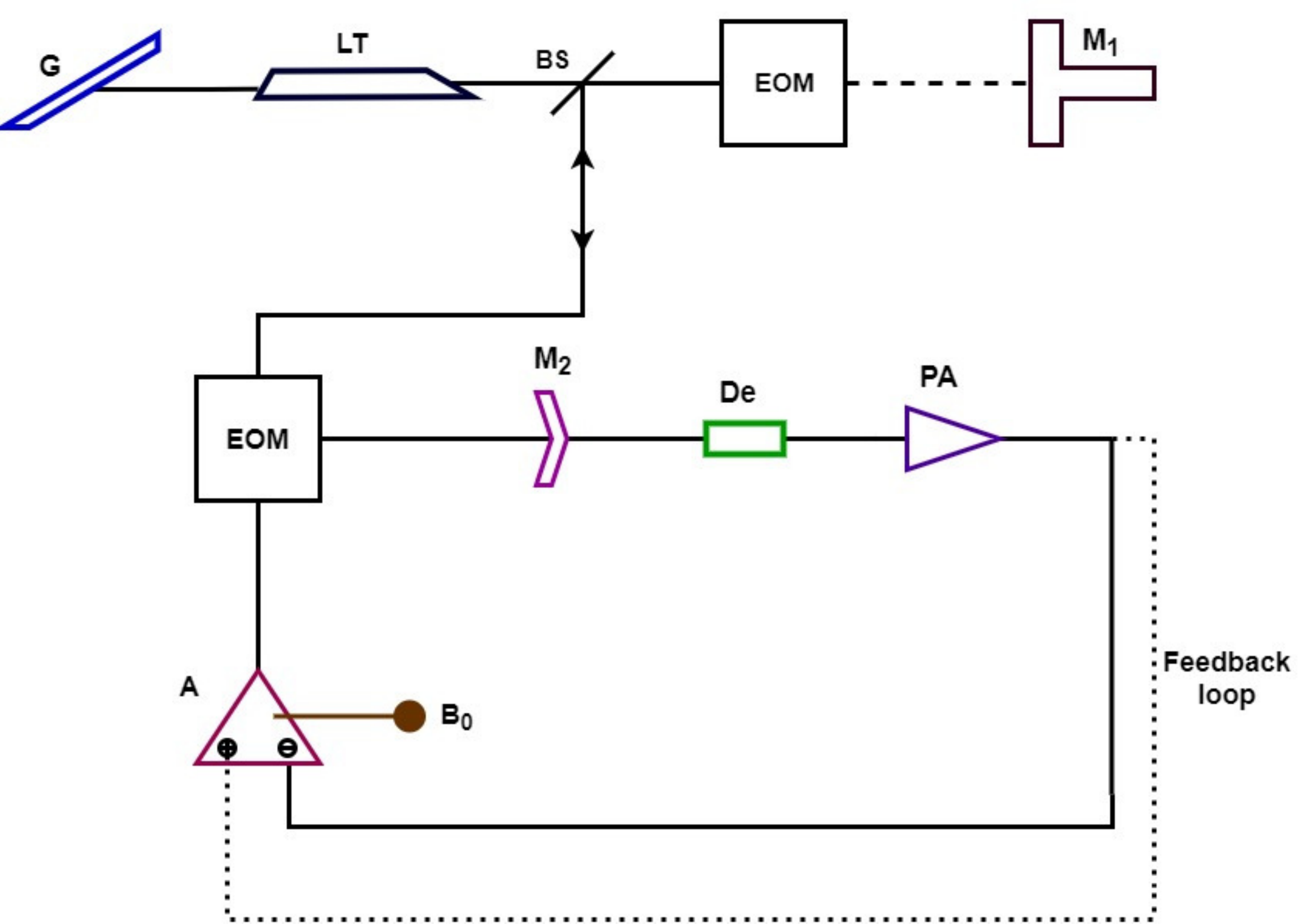}
        \caption{A schematic diagram of a CO$_2$ laser model:   G, diffraction grating;  LT, laser tube; BS, beam splitter;  EOM, electro-optical modulator;  M$_1$, Piezoelectric (PZT) mirror;  M$_2$, out coupling mirror;   De,  HgCdTe detector; PA, pre-amplifier; A, differential amplifier, B$_0$, bias voltage input. The dotted line represents an additional negative feedback which may be used to control chaos of the system.}
        \label{fig:fig1-diagram}
\end{center}
\end{figure*} 
%%%%%%%%%%%%%%%
\par 
In Eq. \eqref{eq-basic-dim},  the modulated injected beam is modeled by the term  $-k_0I\epsilon\cos(\Omega t)$ in which $\epsilon$ is the modulation depth and  $\Omega$ is the driving frequency of modulation. 
By disregarding the term $\propto k_1$ and the equation for $V$ [the third equation in Eq. \eqref{eq-basic-dim}], one can recover the same model as in Ref. \cite{perez1994}.  Also, in absence of the time-dependent perturbation $\propto \epsilon$, one can reproduce a similar model as in Ref. \cite{meucci1997}. Note that  both the time-dependent perturbation and the negative feedback driven by $V$   have been separately used to control chaos in different investigations  \cite{perez1994,meucci1997}. In Ref. \cite{perez1994},  the intesity decay rate was considered to be a constant, however, in Ref. \cite{meucci1997}, the same was modified to involve  a constant decay rate plus a voltage dependent perturbation part.   The motivation of this work is to propose a modified dynamical system for CO$_2$ lasers by  including both these feedback beams, and to study their interplay and  roles for    the onset of chaos and hyperchaos. Furthermore, we also construct a network with the hyperchaotic CO$_2$ lasers as its nodes and study their  synchronization.    The key parameters for generating chaos and hyperchaos are  $\epsilon$ and $B$.  However, the absence of any one of them   results into chaos instead of  hyperchaos   in the system. 
\par 
 A schematic diagram of our model for a possible experimental setup is shown in Fig. \ref{fig:fig1-diagram}.  In this configuration, the leaser beam can be modulated by using (i) a CdTe electro-optic modulator (EOM)  and (ii) an intracavity EOM, and can be fed back to the EOMs by   a mirror ($M_1$)  mounted on a piezoelectric transducer (PZT) and an out coupling mirror ($M_2$). The intensity of the laser beam can be measured using HgCdTe photodiode detector (De) followed by an amplifier (PA).  The CO$_2$ laser may be driven into chaos by injecting  any  one of the modulated feedback beams mentioned above. However, an additional negative feedback, to be obtained from subharmonic components of the laser intensity, may be used to establish hyperchaos in the system \cite{meucci1997}. 
\par  
It is useful to define/redefine the dimensionless variables as
$x=\alpha I/k_0,~y=\alpha k_0/ N,~z= \pi(V-V_0)/V_{\lambda},~n={\alpha N_0}/{k_0},~b= {\pi(V_0-B)}/{V_{\lambda}},~r= {\pi k_0R}/{\alpha V_{\lambda}},~ \zeta=\eta {k_0}/{\alpha},~ \tau=\gamma t$. Also, we set    $l_1={k_0}/{\gamma},~l_2={\Omega}/{\gamma},~l_3={k}/{\gamma}$ and $l_4={\beta}/{\gamma}$. Thus, equations in \eqref{eq-basic-dim} reduce to
\begin{equation}
\begin{split}
\frac{dx}{d\tau}&=x\left[l_1\left(y-1-k_1\sin^2 {z}\right)- l_3\epsilon \cos(l_2\tau)\right] \\
\frac{dy}{d\tau}&=-(y-n)-2l_1xy\\
\frac{dz}{d\tau}&=-l_4\left(z-b+\frac{rx}{1+\zeta x}\right). \label{eq-basic-dimless}
\end{split}
\end{equation}
%\subsection{Dynamical properties} 
\par
We   study the dynamical properties CO$_2$ lasers given by Eq. \eqref{eq-basic-dimless}, and show that the system indeed exhibits chaos and hyperchaos by the control parameters $\epsilon$ and $b$ associated with the modulation depth and the bias voltage. To this end,  we first find the equilibrium points of the system which can be obtained by equating the right-hand sides of  Eq. \eqref{eq-basic-dimless} to zero and finding solutions for $ x,~y$ and $z$. Thus,  an equilibrium  point is obtained as  $P(0,n,b)$ $\forall\tau$. The stability of the system \eqref{eq-basic-dimless} about the fixed point $P$ can now be studied. So, we consider the following perturbed system of equations for $X=(x_1,y_1,z_1)$, where $x_1=x$, $y_1=y-n$ and $z_1=z-b$.  
\begin{equation}
\begin{split}
\frac{dx_1}{d\tau}&=x_1\left[l_1\left(y_1+n-1-k_1\sin^2 (z_1+b)\right) - l_3\epsilon \cos(l_2\tau)\right],\\
\frac{dy_1}{d\tau}&=-y_1-2l_1x_1(y_1+n),\\
\frac{dz_1}{d\tau}&=-l_4\left(z_1+\frac{rx_1}{1+\zeta x_1}\right). \label{eq-model-perturb}
\end{split}
\end{equation}
Equation \eqref{eq-model-perturb} can be rewritten as 
\begin{equation}
\dot{X}=A(X,\tau)X, \label{eq-model-perturb-matrix}
\end{equation}
where $A$ is the coefficient matrix, given by,
\begin{equation}
 A=\begin{bmatrix}
 \phi_1 & 0 & 0 \\
-2l_1(y_1+n) & -1 & 0\\
-{rl_4}/({1+\zeta x_1})&0&-l_4\\
\end{bmatrix}.
\end{equation}
Here, $\phi_i=l_1\left(y_i+n-1-k_1\sin^2(z_i+b)\right) - l_3\epsilon \cos(l_2\tau)$ with $i=1$.
 Next, we assume that Eq. \eqref{eq-model-perturb-matrix} has a solution of the form $X=\Psi(\tau) X(\tau_0)$ where $X(\tau_0)$ denotes the initial value of $X=(x_1,y_1,z_1)$  at $\tau=\tau_0$  and 
 \begin{equation}
 \Psi(\tau)= I+ \int_{\tau_0}^{\tau} A(s)ds +\int_{\tau_0}^\tau A(r)dr\int_{\tau_0}^{r}A(s)ds+\cdots,
 \end{equation}
  i.e.,   $\Psi(\tau)$ is a propagator from the state time $\tau_0$ to $\tau$ and   is equivalently  
   \begin{equation}
   \Psi(\tau) =\lim_{\delta \tau\rightarrow 0} \Pi_{j=1}^n e^{A(\tau_j)\delta \tau},
   \end{equation} 
    where $\tau_0=\tau_1<\tau_2<\tau_3<\cdots<\tau_{n-1}<\tau_n=\tau$, and $\tau=\tau_0 +n\delta \tau$ and $\Psi(\tau_0)=I$, the identity matrix of order $3$.  The matrix $\Psi$  describes how a small change in $\tau$  develops from the initial    state  $X(\tau_0)$   to the final state $X(\tau)$, and it satisfies the following equation. 
\begin{equation}
\frac{d\Psi}{d\tau}=A(\tau)\Psi. 
\end{equation}
The stability of  the system  [Eq. \eqref{eq-model-perturb} or Eq. \eqref{eq-model-perturb-matrix}] can be studied by  finding the eigenvalues (Lyapunov exponents) of the matrix, given by, 
\begin{equation}
\Lambda = \lim_{\tau\rightarrow \infty }\frac{1}{2\tau}{\log(\Psi^{\text T}\Psi)},
\end{equation} 
 where the superscript `T' denotes the transpose of a matrix.  
 \par
\begin{figure*}[h!]
\begin{center}
            \includegraphics[width=6in,height=2.5in]{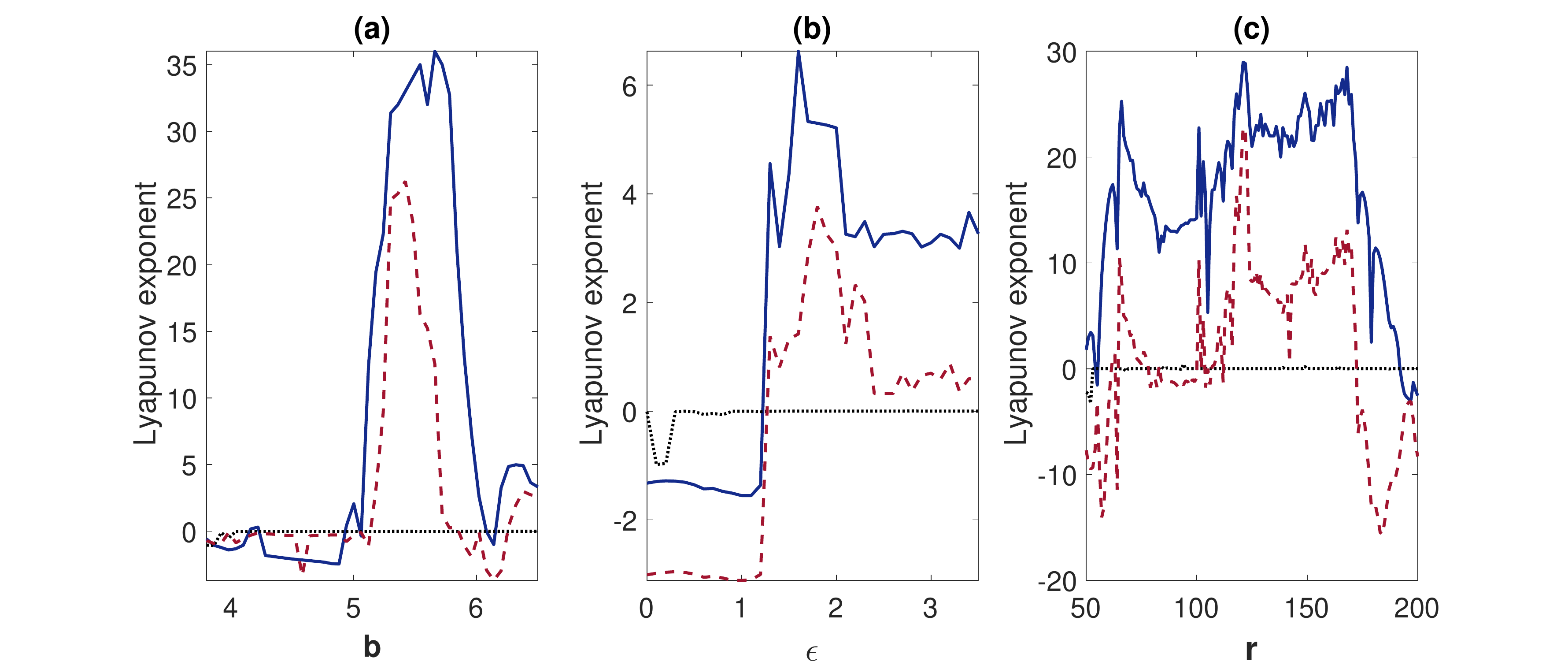}
        \caption{The Lyapunov exponents are shown against the variation of the  parameters $b$, $\epsilon$  and $r$. The  other parameter values corresponding to the subplots (a), (b) and (c), respectively, are  $\epsilon=1.2,~r=200$; $b=4.5,~r=200$  and $\epsilon=1.2,~b=4$.  }
        \label{fig:fig2-lyapunov}
        \end{center}
\end{figure*}
\begin{figure*}[h!]
\begin{center}
            \includegraphics[width=6.0in,height=2.5in]{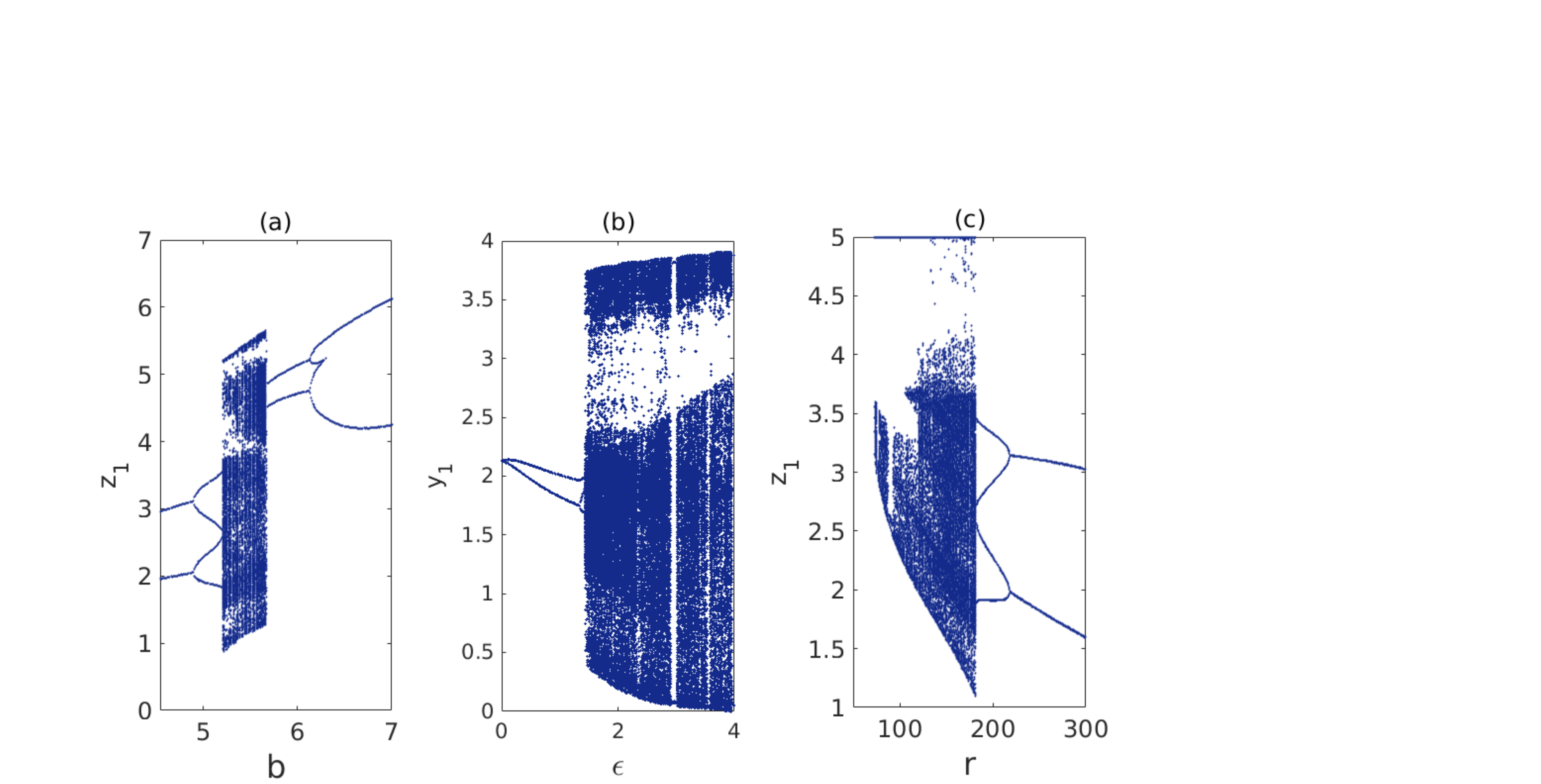}
        \caption{ The bifurcation diagrams are shown with respect to the   parameters $b$, $\epsilon$  and $r$.   The  other parameter values corresponding to the subplots (a), (b) and (c)  are the same as   in Fig. \ref{fig:fig2-lyapunov}.  }
  \label{fig:bifurcation}      
\end{center}
\end{figure*}

    The Lyapunov exponents are  shown in Fig. \ref{fig:fig2-lyapunov} with the variations of the parameters $b$, $\epsilon$ and $r$ for some fixed values of other parameters, i.e., $n=4,~l_1=28.57,~l_2=30,~l_3=25,~l_4=6.452,~k_1=2$ and $\zeta=25.57$.    
%%%%%%%%%%%%%%%%%%%%%%%%%%%
%%%%%%%%%%%%%%%%%%%%%%%%%%%%%%%%%
From Fig. \ref{fig:fig2-lyapunov}, we find that there exist   different ranges of values of the parameters in which   the  system may exhibit   periodic, chaotic or hyperchaotic states.  In Fig. \ref{fig:fig2-lyapunov}, subplot (a) shows that for some fixed values of  $\epsilon=1.2$ and $r=200$, and others remain as above, there is a wide range of values of $b$ for which   the two Lyapunov exponents remain  positive and  another one  remains negative. In this case, the periodicity occurs for $0<b<5$, and the chaotic or hyperchaotic states may occur  either in  $5\lesssim b\lesssim6.5$ or $b>6.5$.    However, from the subplot (b)  one can predict that while the periodicity may occur in the range $0<\epsilon\lesssim1.5$, the hyperchaotic state is more likely to occur  for  $\epsilon \gtrsim1.43$ with some fixed values of $b=4.5$ and $r=200$ as at least two Lyaponv exponents assume positive values therein.  This means that the injected modulated feedback beam ($\sim\cos\Omega t$) or that associated with the bias voltage $B$ is the prerequisite   for the onset of hyperchaos in CO$_2$ lasers. Furthermore, it is also noticed that the system with chaotic/hyperchaotic states reaches towards a steady state as the value of $r$ increases [subplot (c)].  In order to justify the results of Fig. \ref{fig:fig2-lyapunov}, we also show the bifurcation diagrams corresponding to the parameters $b$, $\epsilon$ and $r$ as in  Fig. \ref{fig:bifurcation}.  From the subplot (a), it is clear that there are some ranges of values of $b$ for which the periodicity and chaotic or hyperchaotic states  occur one after another. Subplot (b)  confirms that the periodicity route to chaos/hyperchaos occurs in some ranges of values of $\epsilon$.  Furthermore, the fact that the higher values of $r$ leads to a steady state   is evident from the subplot (c).  
%%%%%%%%%%%%%%%%%%%%%%%%
%%%%%%%%%%%%%%%%%%%%%%%%%
\par
\begin{figure*}[h!]
\begin{center}
            \includegraphics[width=6in,height=2.5in]{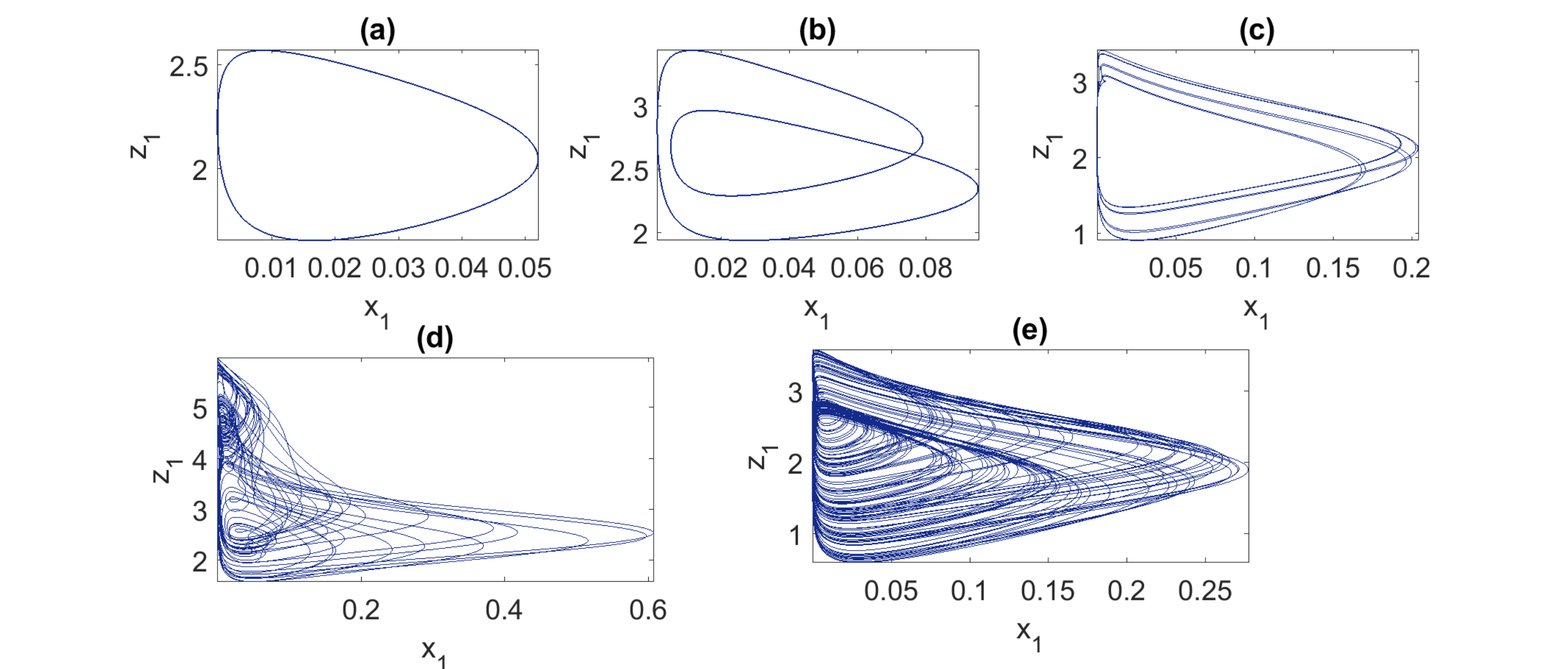}
        \caption{Different phase portraits are shown with the variation of the parameter $\epsilon$:   (a) stable and single periodic orbit   $(\epsilon=1.2)$   (b) stable and double periodic orbit $(\epsilon=1.4)$  (c) unstable and multi-periodic orbit $(\epsilon=1.54)$ (d) chaotic orbit $(\epsilon=1.7)$ and (e) hyperchaotic orbit $(\epsilon=2)$.  The other parameter values are  
         $b=4.5,~r=200$,   $n=4,~l_1=28.57,~l_2=30,~l_3=25,~l_4=6.452,~k_1=2$ and $\zeta=25.57$. }
  \label{fig:phase-portraits}      
\end{center}
\end{figure*}

 
 For an illustration purpose, we plot different phase portraits (Fig. \ref{fig:phase-portraits}) to show that the periodic [subplots (a) and (b)], multi-periodic [subplot (c)], chaotic [subplot (d)] and hyperchaotic [subplot (e)] states of the system coexist for different ranges of values of the parameters, especially   $b,~\epsilon$ and $r$. 
%%%%%%%%%%%%%%%%%%%%%%%
%%%%%%%%%%%%%%%%
\section{Network of coupled  hyperchaotic CO$_2$ lasers and synchronization}
 We construct a network graph in which each edge of the graph is connected to a finite number of nodes. We choose the hyperchaotic CO$_2$ lasers, given by Eq. \eqref{eq-basic-dimless}, as the nodes or oscillators, and couple  them  through the $x$ and $y$ variables. Here, the coupling with  the  $z$ variable is not introduced  as it  corresponds to the voltage  obtained through a feedback loop. Next, we derive a variational equation from the coupled set of equations and  study the stability   of the synchronous state of the coupled oscillators using the method of MSF.  A numerical simulation approach is also performed to find the synchronization errors and to validate the theory of MSF for a finite number of nodes. 
\par 
 The network of $n$ identical oscillators  that are linearly coupled are given by

%%%%%%%%%%%%%%%%
\begin{equation}
\begin{split}
\frac{dx_i}{d\tau}&=x_i\left[l_1\left(y_i-1-k_1\sin^2 {z_i}\right)\right] - l_3\epsilon \cos(l_2\tau)\\
&+\sigma_1\sum_{j=1}^nG_{ij}x_j,\\
\frac{dy_i}{d\tau}&=-(y_i-n)-2l_1x_iy_i+\sigma_2\sum_{j=1}^nG_{ij}y_j,\\
\frac{dz_i}{d\tau}&=-l_4\left(z_i-b+\frac{rx_i}{1+\zeta x_i}\right), \label{eq-network}
\end{split}
\end{equation}
%%%%%%%%%%%%%%
which can be rewritten  in the following form. 
\begin{equation}
\dot{X}_i=F(X_i)+\sigma\sum_jG_{ij}X_j. \label{eq-network1}
\end{equation}
Here,  $X_i=(x_i,y_i,z_i)$   denotes the three-dimensional vector of the  dynamical variables (corresponding to $x,~y$ and $z$) of the $i$th node,   $\sigma=(\sigma_1,\sigma_2,0)$ is the coupling strength, and $G=(G_{ij})_{n\times n}$ is a Laplacian matrix of order $n$ with zero row-sums (i.e., $\sum_jG_{ij}=0$ such that the synchronization manifold, defined by the $n-1$ constraints $X_1=X_2=\cdots=X_n$, is an invariant manifold)  and non-negative off diagonal elements.  For different choice of the matrix $G$, one can have  different connectivity of nodes. For example, $G=G_1$ or $G_2$ according to when one considers the  RDC   or SC networks where \cite{pecora1998,belykh2004}
\begin{equation} 
 G_1=\begin{bmatrix}
-2&1&0&0&...&1\\
1&-2&1&0&...&0\\
0&1&-2&1&...&0\\
...&...&...&...&...&...\\
1&0&0&...&1&-2
\end{bmatrix} \label{eq-matrix-G1}
\end{equation}  
 and 
 \begin{equation} 
G_2=\begin{bmatrix}
-(n-1)&1&1&1&...&1\\
1&-1&0&0&...&0\\
1&0&-1&0&...&0\\
...&...&...&...&...&...\\
1&0&0&...&0&-1
\end{bmatrix}. \label{eq-matrix-G2}
\end{equation}   
Next, we combine the matrices $G_1$ and $G_2$   with the coupling strengths   $\sigma_1$ and   $\sigma_2$ to obtain the corresponding matrices $H=H_1$ and $H=H_2$   which contain  $3\times 3$ matrix blocks, so that     Eq. \eqref{eq-network1} can be recast as
\begin{equation}
\dot{X_i}=F(X_i)+\sum_{j}H_{ij}X_j,  \label{eq-network2}
\end{equation}. 
where $H=(H_{ij})_{n\times n}$. Thus, for $n$ nodes or oscillators   we get the matrices $H_1$ and $H_2$ of order  $3n$:
%%%%%%%%%%%%%%%%%%%%%%%%%%%%%%%%
%%%%%%%%%%%%%%%%%%%%%%%%%%%%
\begin{equation}
 H_{1}=\begin{bmatrix}
-2J_3 & J_3 & O_3 &...&O_3 & J_3\\
J_3 & -2J_3 & J_3 &...&O_3 & O_3\\
O_3 & J_3 & -2J_3 &...&O_3 & O_3\\
... & ... & ... &  ...&... &...\\
... & ... & ... &  ...&... &...\\
-2J_3&O_3 & O_3 & ...& J_3 &-2J_3
\end{bmatrix},
\end{equation}
  and 
 \begin{equation}
H_{2}=\begin{bmatrix}
(-n+1)J_3 & J_3 & J_3 &.....&J_3 & J_3\\
J_3 & -J_3 & O_3 &.....&O_3 & O_3\\
J_3 & O_3 & -J_3 &.....&O_3 & O_3\\
... & ... & ... &  .....&... &....\\
... & ... & ... &  .....&... &....\\
-J_3&O_3 & O_3 & ..... & O_3 &-J_3
\end{bmatrix}
\end{equation}
 with $O_3$ denoting the null matrix of order $3$ and
 \begin{equation}
J_3=\begin{bmatrix}
\sigma_1&0&0\\
0&\sigma_2&0\\
0&0&0
\end{bmatrix}.
%~~O_3=\begin{bmatrix}
%0&0&0\\
%0&0&0\\
%0&0&0
%\end{bmatrix}.
\end{equation}
%%%%%%%%%%%%%%%%%%%
In order to study the  synchronization of $n$ coupled oscillators, given by Eq. \eqref{eq-network2} and its stability, we decompose them into driving and response systems, i.e., we assume the driving system as 
\begin{equation}
\begin{split}
\dot{x_i}=&f_{1}(x_i,y_i,z_i),\\
 \dot{y_i}=&f_{2}(x_i,y_i,z_i), \\ 
 \dot{z_i}=&f_{3}(x_i,y_i,z_i),
 \end{split}
 \end{equation}
 and the response system as
 \begin{equation}
\begin{split}
\dot{x_j}=&f_{1}(x_j,y_j,z_j),\\
 \dot{y_j}=&f_{2}(x_j,y_j,z_j), \\ 
 \dot{z_j}=&f_{3}(x_j,y_j,z_j),
 \end{split}
 \end{equation}
  where $i,j=1,2,...,n$ and $i\neq j$, and the functions $f_{1},~f_{2}$ and $f_{3}$ represent  the right-hand sides of  Eq. \eqref{eq-network}. Here, 
we note that for the $z$ variable, no coupling strength is considered for the reason mentioned before. Also,  in the process of synchronization,  we must have $\parallel (x_{i+1},y_{i+1},z_{i+1})-(x_i,y_i,z_i)\parallel \longrightarrow 0$ as $t\longrightarrow \infty$.  So, synchronization of   $n$ identical nodes, given by Eq.  \eqref{eq-network2}, can be studied by considering the  evolution of the variations $\xi_i^1=x_{i+1}-x_i, ~ \xi_i^2=y_{i+1}-y_i,~\xi_i^3=z_{i+1}-z_i=0$ for  $i=1,2,...,n-1$ and $\xi_n^1=x_n-x_{1},~ \xi_n^2=y_n-y_1,~\xi_n^3=z_n-z_1$. Thus, 
\begin{equation}
\begin{split}
\dot{\xi}_i^1=&f_1(x_{i+1},y_{i+1},z_{i+1})-f_1(x_i,y_i,z_i),\\
\dot{\xi}_i^2 =&f_2(x_{i+1},y_{i+1},z_{i+1})-f_2(x_i,y_i,z_i),\\
\dot{\xi}_i^3 =&0,
\end{split}
\end{equation}
 or,
 \begin{equation}
\begin{split}
 \dot{\xi}_i^1 =&f_1(x_{i}+\xi_i^1,y_{i}+\xi_i^2,z_{i})-f_1(x_i,y_i,z_i),\\
 \dot{\xi}_i^2 =&f_2(x_{i}+\xi_i^1,y_{i}+\xi_i^2,z_{i})-f_2(x_i,y_i,z_i),\\
 \dot{\xi}_i^3 =&0.
 \end{split}
\end{equation}
 In what follows, we Taylor expand  the functions $f_1$ and $f_2$  about $(x_i,y_i,z_i)$ and keep  terms up to the first orders of $\xi_i^1$ and $\xi_i^2$  to obtain for $i=1,2,...,n$ the following equation.
%\begin{equation}
%\begin{split}
% \dot{\xi_i^1}=&(\frac{\delta f_1}{\delta x} ~~ \frac{\delta f_1}{\delta y})\times(\xi_i^1 ~~ \xi_i^2)^T,\\
%  \dot{\xi_i^2}=&(\frac{\delta f_2}{\delta x} ~~ \frac{\delta f_2}{\delta y})\times(\xi_i^1 ~~ \xi_i^2)^T, 
% \end{split}
%\end{equation}
\begin{equation}
 \begin{bmatrix}
 \dot{\xi}_i^1 \\
 \dot{\xi}_i^2\\
  \dot{\xi}_i^3
 \end{bmatrix}= \begin{bmatrix}
 \frac{\delta f_1}{\delta x} & \frac{\delta f_1}{\delta y}&0\\
 \frac{\delta f_2}{\delta x} & \frac{\delta f_2}{\delta y}&0\\
 0&0&0
 \end{bmatrix} \begin{bmatrix}
 \xi_i^1\\
 \xi_i^2\\
 \xi_i^3
 \end{bmatrix}.  
 \end{equation}
Thus,  a variational equation of Eq. \eqref{eq-network2} is obtained as  
  \begin{equation}
   \dot{\xi}=(F+H\otimes E)\xi, \label{eq-variation}
  \end{equation}
  where  $\xi=(\xi_1^1,\xi_1^2,\xi_1^3;\xi_2^1,\xi_2^2,\xi_2^3,...,\xi_n^1,\xi_n^2,\xi_n^3)$, $E$ and $F$ are the   matrices (each of order $3n$), given by,
   \begin{equation}
 E=\begin{bmatrix}
  P&O_3&O_3&...&O_3\\
  O_3&P&O_3&...&O_3\\
  ...&...&...&...&...\\
  O_3&O_3&O_3&...&P
  \end{bmatrix},~F=\begin{bmatrix}
  F_1&O_3&O_3&...&O_3\\
  O_3&F_2&O_3&...&O_3\\
  ...&...&...&...&...\\
  O_3&O_3&O_3&...&F_n
  \end{bmatrix},
  \end{equation}
      with 
  \begin{equation}
 P=\begin{bmatrix}
  1&0&0\\
  0&1&0\\
  0&0&0
  \end{bmatrix},~
   F_i=\begin{bmatrix}
 \phi_i & l_1x_i&0\\
  -2l_1(y_i+n) & -1-2l_1x_i&0\\
  0&0&0 
  \end{bmatrix}  
    \end{equation}     
 for $i=1,2,...,n$. Also, in Eq. \eqref{eq-variation}, $H$ is $H_1$ or $H_2$ according to when we choose  RDC or SC  networks.
\subsection{Stability analysis of synchronous states} \label{sec-sub-stab-ana-synch-state}
The master stability function,  which relies on finding the maximum Lyapunov exponent for the least stable transverse mode of the synchronization manifold along with the eigenvalues of the connectivity matrix, has been recognized as   one of the most powerful tools for determining  the stability of  synchronous states of linearly coupled oscillators \cite{pecora1998}. Such MSF allows one to quickly establish whether any linear coupling of oscillators will result into a stable synchronous dynamics. It also reveals which desynchronization bifurcation mode will occur due to changes of the coupling scheme or strength.  Furthermore,  it simplifies a large-scale networking  system to a node-size system via diagonalization and decoupling  as long as the inner coupling functions for all node pairs are identical. 
\par 
 We recall the  variational equation   \eqref{eq-variation}, which will be used to calculate the Lyapunov exponents, as
 \begin{equation}
   \dot{\xi}=(F+H\otimes E)\xi, \label{eq-variation-1}
  \end{equation}
 We note that in Eq. \eqref{eq-variation-1} the matrix $F$ is a block diagonal matrix which involves $3\times3$ block matrices $F_i$. The matrix $H\otimes E$ can be treated by diagonalizing  the symmetric matrix $H$ (i.e., $H_1$ or $H_2$) as $E$ is already a diagonal matrix. Thus, a block diagonalized variational equation can be formed with each block having the form
  \begin{equation}
  \dot{\xi}_k=(F_k+\lambda_k P)\xi_k, \label{eq-variation-2}
\end{equation} 
where $\lambda_k$ is an eigenvalue of a block diagonal $3\times3$ matrix element of $H$ with $k=0,1,2$. There are, in fact,  $n$ equations of the form of  Eq. \eqref{eq-variation-2} and,    in general, the eigenvalues $\lambda_k$ for each   diagonal $3\times3$ block matrix element of $H$ will be different from the other blocks.   The equation with $k=0$ where $\lambda_0=0$  corresponds to the variational equation for the synchronization manifold defined by $X_1=X_2=\cdots=X_n$. So, there are $3n$ eigenvalues for $n$ numbers of  $3\times3$ blocks of which at least $n$ eigenvalues are zero.     The maxima $\gamma^\text{max}_i, ~i=1,2,...,n$,  of the Lyapunov exponents  are obtained  for each block variational equation \eqref{eq-variation-2}, and   finally, the maximum of these maxima, i.e.,  $\gamma_\text{max}=\text{max}_i\lbrace\gamma^\text{max}_i\rbrace$ ($n$ times)  is obtained as the MSF of the variational equation \eqref{eq-variation-1} with the coupling parameters $\sigma_1$ and $\sigma_2$. If $\gamma_\text{max}<0$,   the synchronous state is said to be stable at the coupling strengths, otherwise, it is unstable for $\gamma_\text{max}>0$.  
%%%%%%%%%%%%%%%%%%%% 
\begin{figure*}[h!]
    \begin{center}
            \includegraphics[width=6in,height=2.5in]{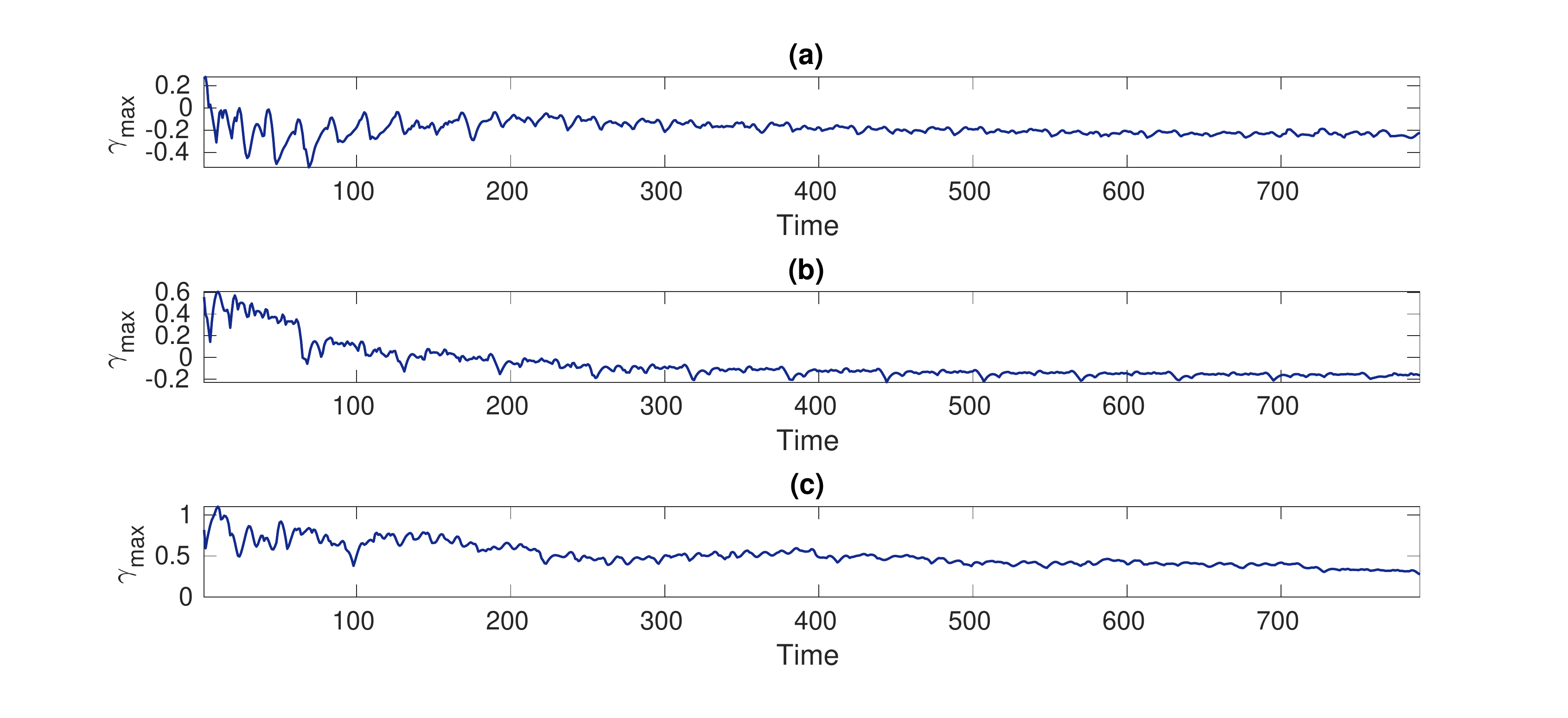}
        \caption{The maximum Lyapunov exponent of the variational equation \eqref{eq-variation-2}   is shown  for nearest-neighbor ring of  diffusive coupling (RDC) networks with different number of nodes: (a) $20$, (b) $24$ and (c) $28$. The coupling parameters are $\sigma_1=28$ and $\sigma_2=26$. }
        \label{fig:lyapunov-max-ring}
\end{center}
\end{figure*}
%%%%%%%%%%%%%%%%%%%%%%%%%% 
%%%%%%%%%%%%%%%%%%%% 
\begin{figure*}[h!]
\begin{center}
            \includegraphics[width=6in,height=2.5in]{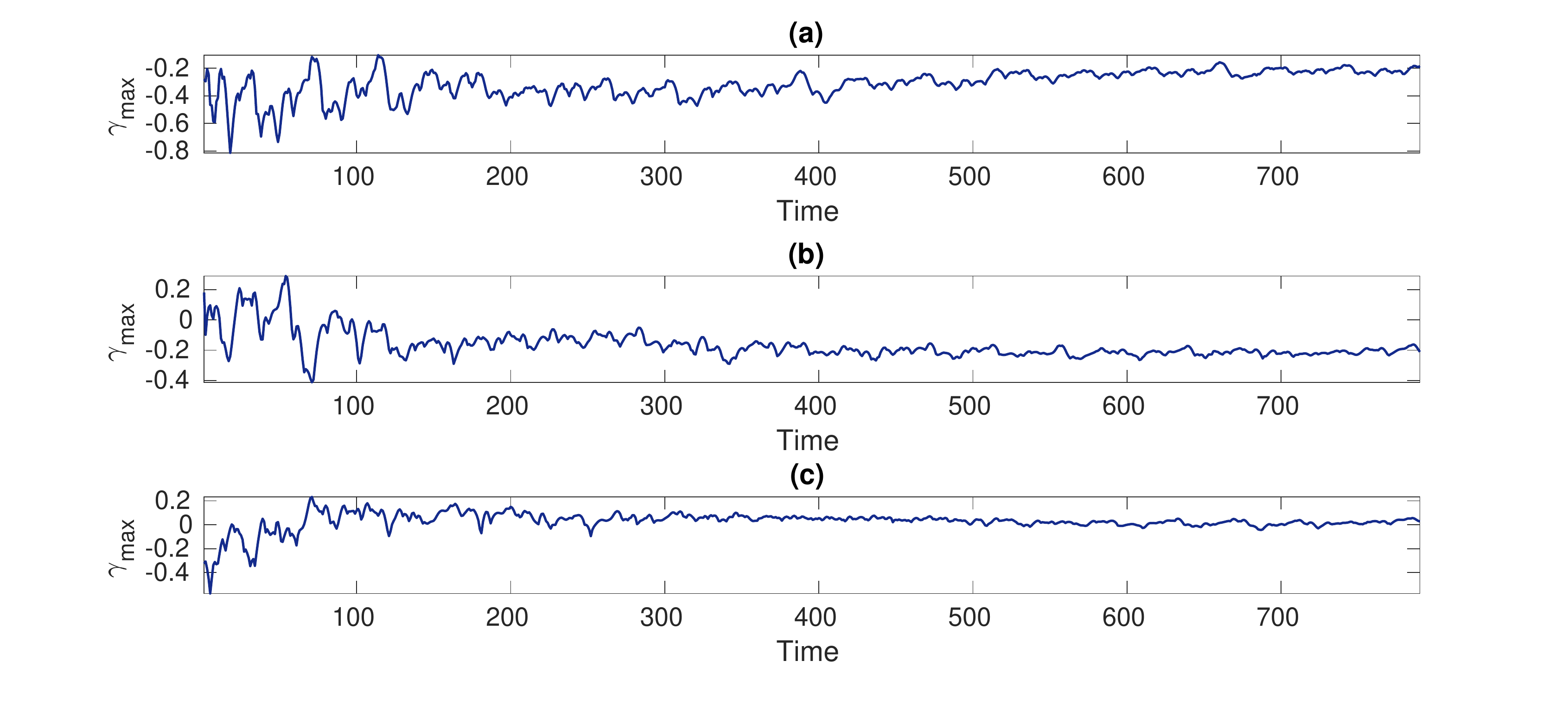}
        \caption{The same as in Fig. \ref{fig:lyapunov-max-ring}, but for   star coupling (SC) networks with (a) $20$, (b) $24$ and (c) $28$ nodes. The coupling parameters are $\sigma_1=5$ and $\sigma_2=3.2$.  }
        \label{fig:lyapunov-max-star}
\end{center}
\end{figure*}
%%%%%%%%%%%%%%%%%%%%%%%%%%

 \par
\begin{figure*}[]
    \centering
    \begin{subfigure}[]%{0.5\textwidth}
        \centering
        \includegraphics[scale=.4]{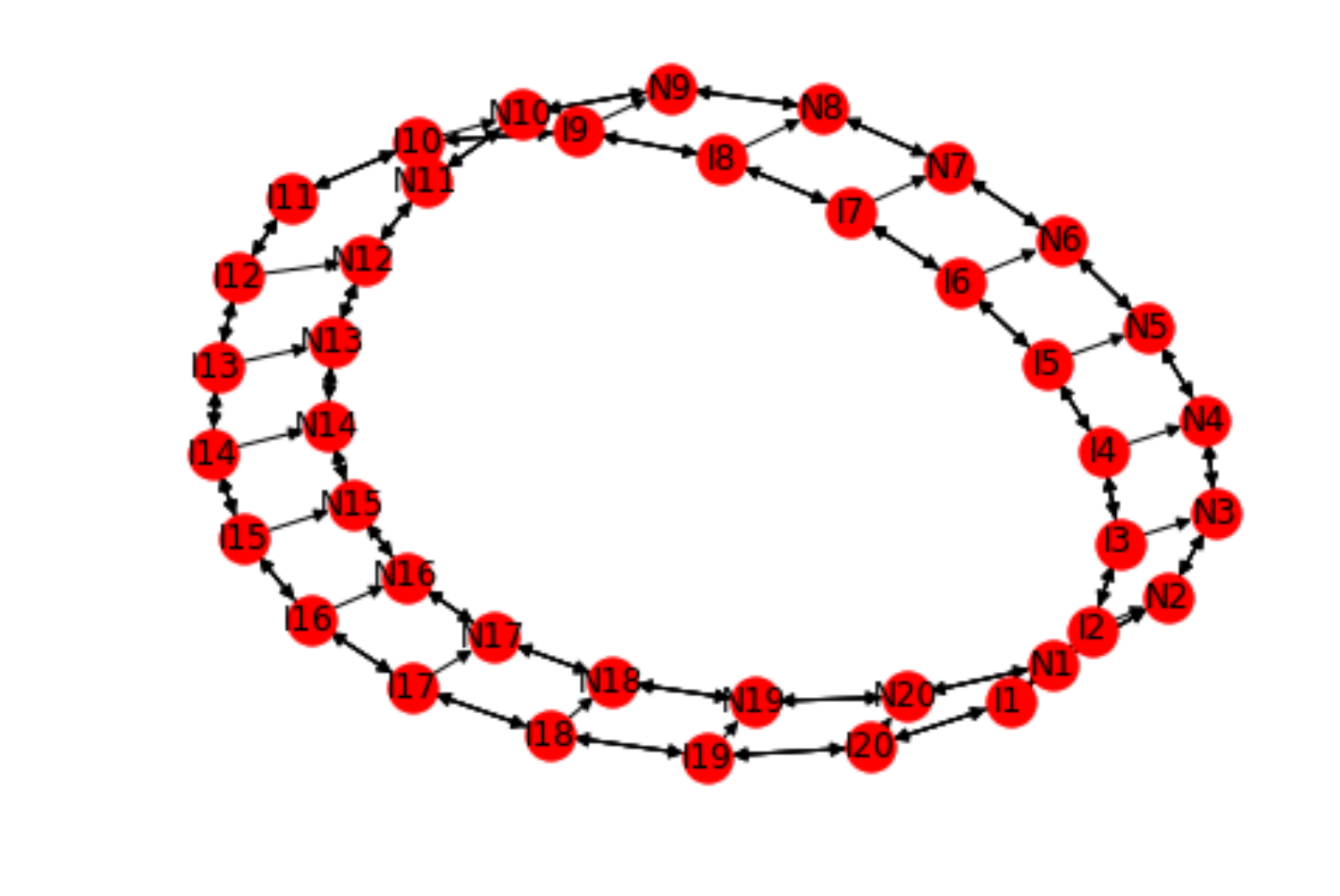}
       %\caption{fig 1}
       \label{fig:ring-coupling}
    \end{subfigure}%
    ~ 
    \begin{subfigure}[]%{0.5\textwidth}
        \centering
        \includegraphics[scale=.4]{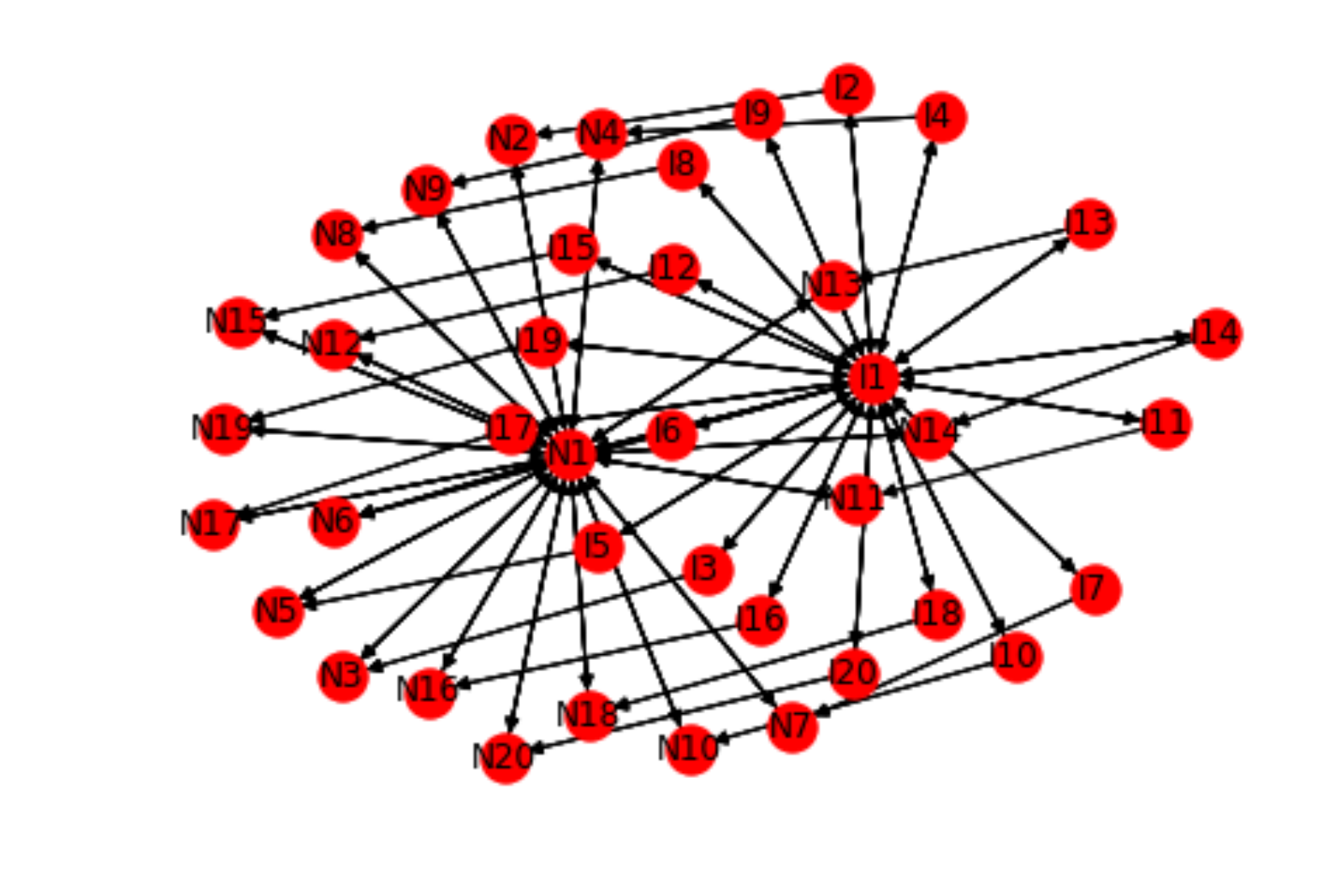}
        %\caption{encrypted green component qqplot}
        \label{fig:star-coupling}
    \end{subfigure}
        
    \caption{A schematic diagram for the synchronization of $20$ nodes  in a  ring of diffusively coupled [subplot (a)] and start coupled [subplot (b)] networking system.}
    \label{fig:diagram-ring-star}
\end{figure*}

 
 We numerically calculate  the Lyapunov exponents of the variational equation \eqref{eq-variation-2}   as functions of $\sigma_1$ and $\sigma_2$ for both the RDC and SC   networks with different number of nodes. The corresponding maxima of these exponents $(\gamma_\text{max})$ are plotted against  time for (a) $20$, (b) $24$ and (c) $28$ nodes as shown in Figs. \ref{fig:lyapunov-max-ring} and \ref{fig:lyapunov-max-star}. We find that given a coupling strength,  the Lyapunov exponent, $\gamma_\text{max}<0$  in a bounded region for a longer time for both types of networks with at most $24$ number of nodes [as $\gamma_\text{max}>0$ for $n>24$, see, e.g., subplot (c) for $n=28$], implying that the synchronous state reaches a steady state at that coupling strength. However, while lower values of $\sigma_1,~\sigma_2~\sim o(1)$ can lead to the stable synchronization for SC networks (Fig. \ref{fig:lyapunov-max-star}), relatively higher coupling strengths  are required for the  RDC networks (Fig. \ref{fig:lyapunov-max-ring}).  A schematic diagram  for the synchronization of both the RDC and SC   networks with $20$ nodes is shown in Fig. \ref{fig:diagram-ring-star}.     
 %%%%%%%%%%%%%%%%%%%%%%%%%%%%%%%%%%%%%%%%%%%%%%%
%%%%%%%%%%%%%%%%%%%%%%%%%%%%%%%%%%%%%%%
%%%%%%%%%%%%%%%%%%% 
\subsection{Synchronization and chimera states} \label{sec-sub-synchro-diff-nodes}
\begin{figure*}[h!]
    \begin{center}
        \includegraphics[width=6in,height=2.5in]{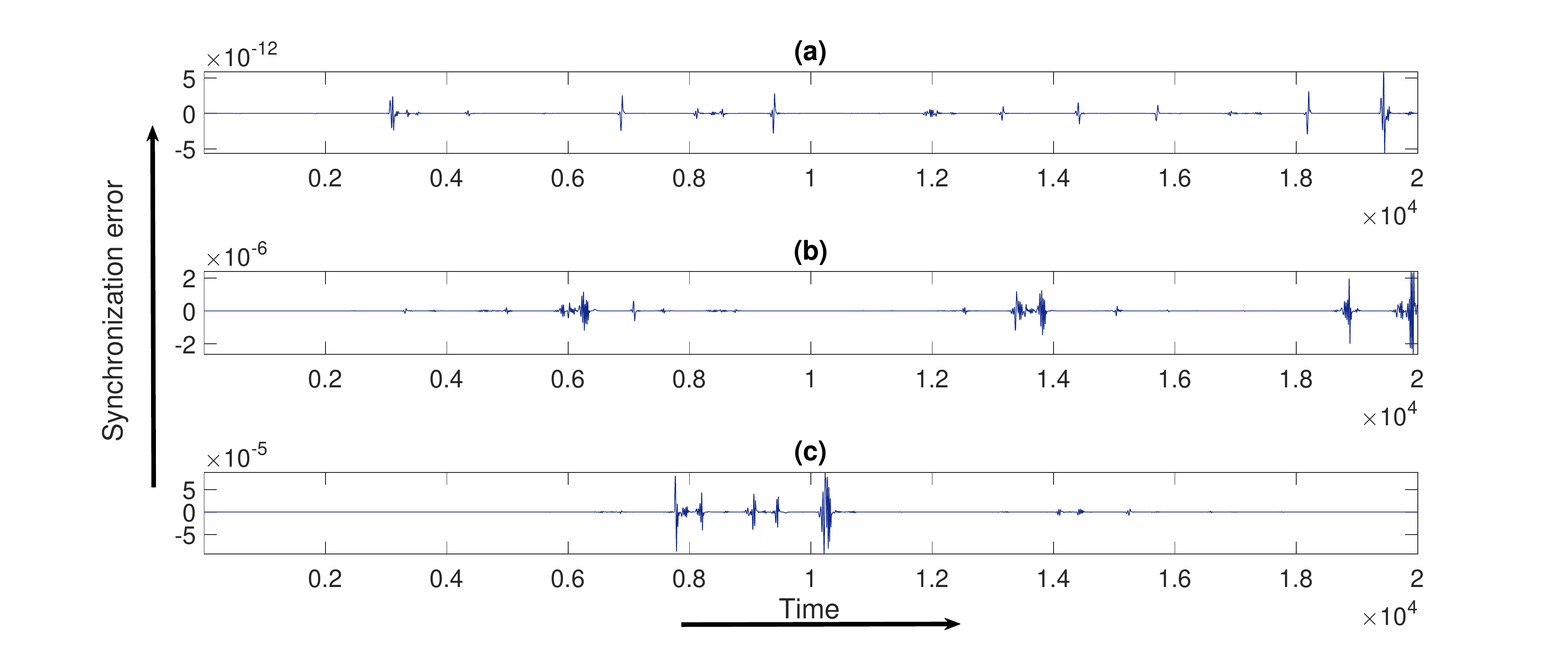}
        \caption{Synchronization error ($||\xi||$) is shown  with respect to time ($\tau$) for (a) $20$   (b)$24$ and (c) $28$ nodes   in a ring coupling network.  The coupling parameters are $\sigma_1=28$ and $\sigma_2=26$.}.
      \label{fig:synchro-error-ring}  
\end{center}
\end{figure*}

\begin{figure*}[h!]
\begin{center}
            \includegraphics[width=6in,height=2.5in]{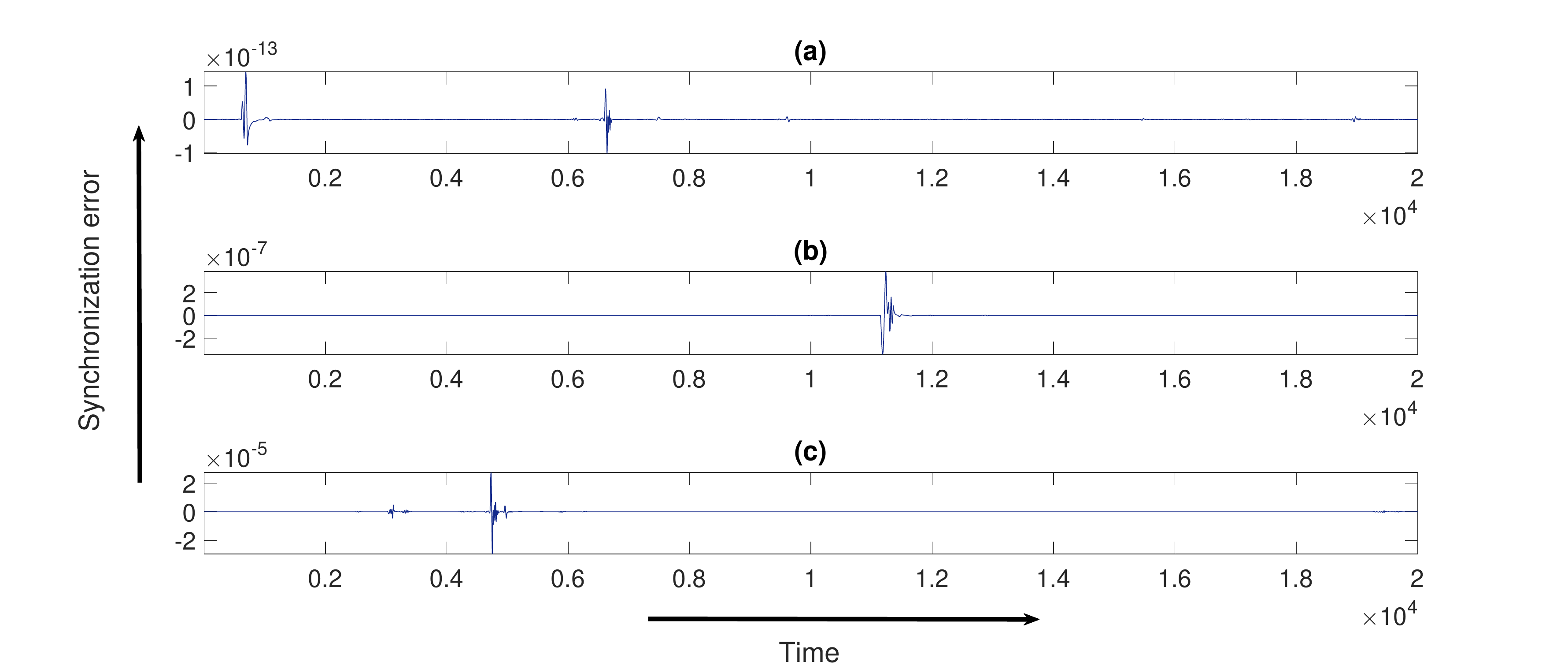}
                    \caption{Synchronization error ($||\xi||$) is shown with respect to time ($\tau$) for (a) $20$   (b)$24$ and (c) $28$ nodes in a star coupling network.    The coupling parameters are $\sigma_1=5$ and $\sigma_2=3.2$}.
       
\label{fig:synchro-error-star}  
\end{center}
\end{figure*}

In order to validate our results in Sec. \ref{sec-sub-stab-ana-synch-state}, we employ the Runge-Kutta scheme to calculate    the synchronization errors corresponding to different sets of (e.g., $n=20$, $24$ and $28$) coupled equations, given by Eq. \eqref{eq-network}, that exhibit hyperchaos.    The simulation results are displayed in Figs.  \ref{fig:synchro-error-ring} and \ref{fig:synchro-error-star}. These basically represent the synchronization errors between the laser intensities of different nodes, i.e., $x_{i+1}-x_i$ at the hyperchaotic states.     We find that the synchronization of the coupled oscillators is achieved through the inclusion of the coupling terms $\propto\sigma$ and  the corresponding errors for the  intensities of the $i+1$-th and $i$-th nodes are   $\lesssim10^{-12}$, $10^{-6}$ or  $10^{-5}$ according to when we choose $n=20$, $24$ or $28$ nodes.   Thus, from the results as in Figs. \ref{fig:lyapunov-max-ring} and \ref{fig:lyapunov-max-star}, and \ref{fig:synchro-error-ring} and \ref{fig:synchro-error-star}, we can conclude that the synchronization in networks of hyperchaotic CO$_2$ lasers can be possible for at most $24$ nodes as evident from the sign of the maximum Lyapunov exponent ,i.e., $\gamma_{\text max}<0$ and the synchronization error  $\lesssim10^{-6}$. 
%%%%%%%%%%%%%%%%%%%%%%%%%%%%%%%%
%%%%%%%%%%%%%%%%%%%%%%%%%%%%%%%%%%%%%%%%%%%%%%%
%%%%%%%%%%%%%%%%%%%%%%%%%%%%%%%%%%%%%%
\par
\begin{figure*}[h!]
    \begin{center}
        \includegraphics[width=6in,height=2.5in]{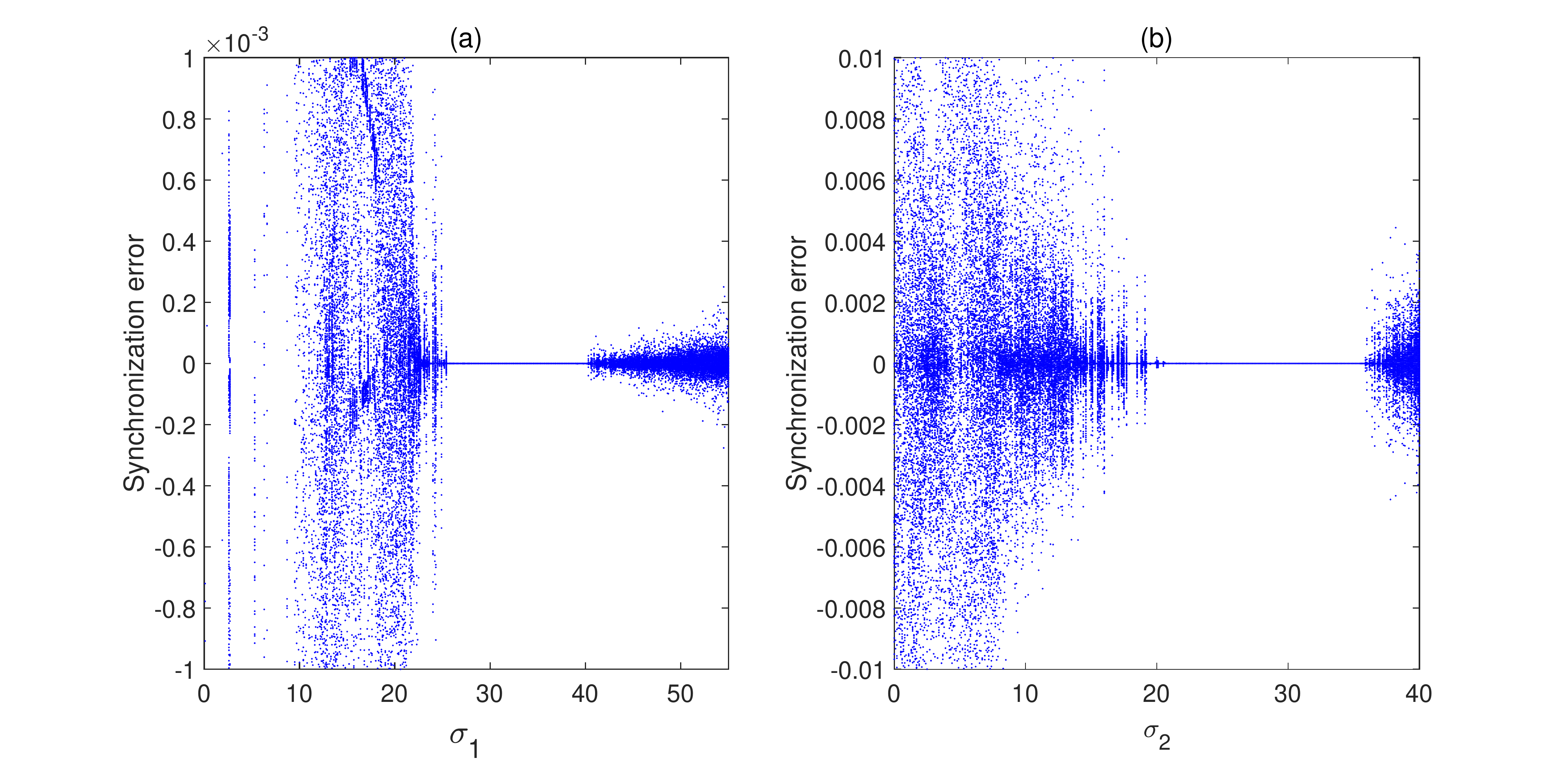}
        \caption{ Synchronization errors ($||\xi||$) are plotted against the coupling   parameters   (a) $\sigma_1$  and (b)$\sigma_2$ to show that the chimera states exist  in a RDC network with $24$ nodes. } 
      \label{fig:synchro-error-ring-chimera}  
\end{center}
\end{figure*}

\begin{figure*}[h!]
\begin{center}
            \includegraphics[width=6in,height=2.5in]{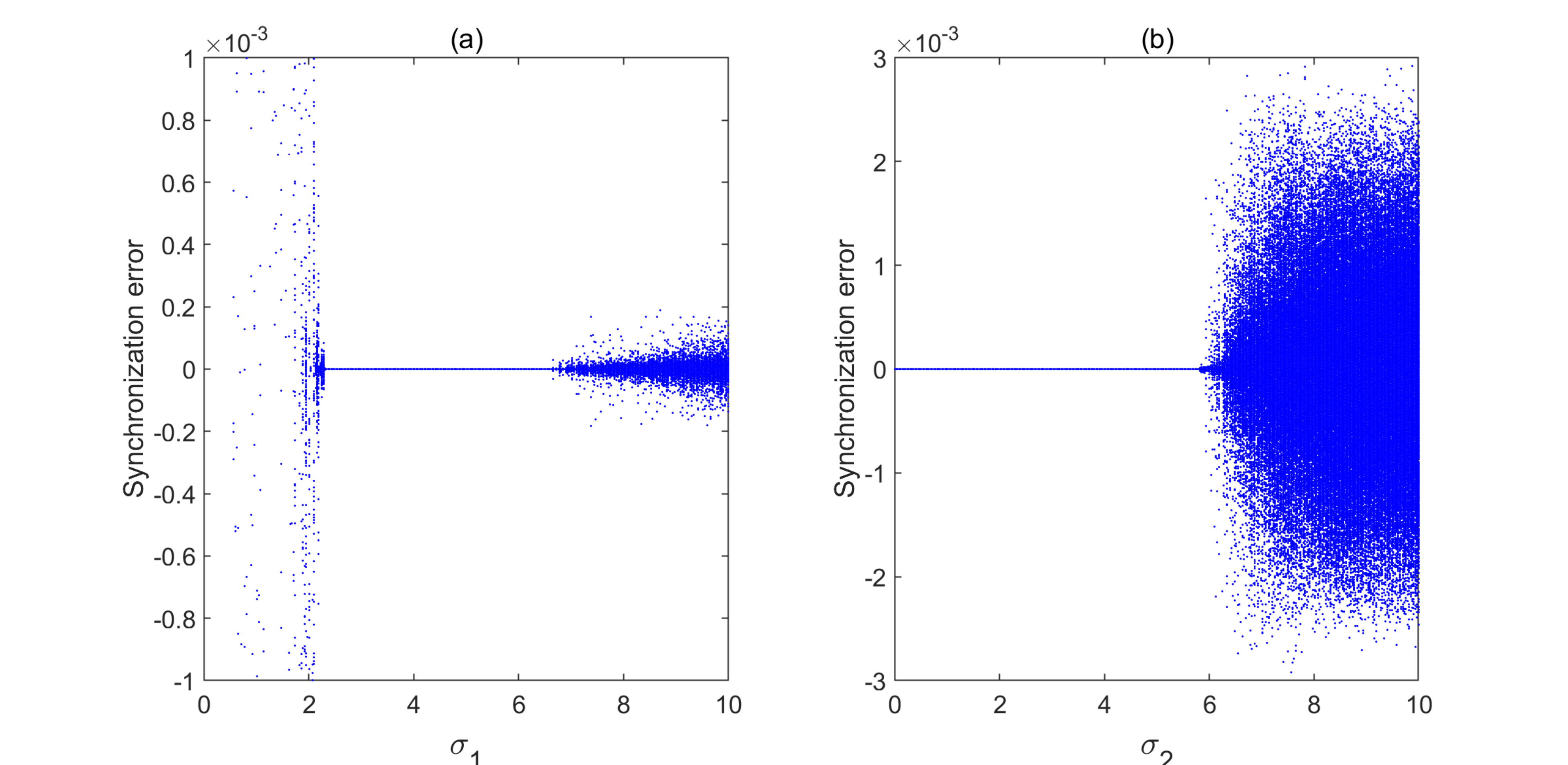}
                    \caption{Synchronization errors ($||\xi||$) are plotted against the coupling parameters   (a) $\sigma_1$  and (b)$\sigma_2$ to show that the chimera states exist   in a SC network with $24$ nodes.} 
       
\label{fig:synchro-error-star-chimera}  
\end{center}
\end{figure*}

 From the characteristics of the Lyapunov exponents (Figs. \ref{fig:lyapunov-max-ring} and \ref{fig:lyapunov-max-star}) and the synchronization errors (Figs. \ref{fig:synchro-error-ring} and \ref{fig:synchro-error-star}) we conclude that the steady states of synchronizations of RDC and SC networks (with synchronization error $\lesssim10^{-6}$) can be reached with at most $24$ number of nodes. However, the chimera states in the networks may coexist  in some time intervals where  the synchronization  does not occur for both the RDC and SC networks.    In order to clarify it,  we plot the synchronization errors against the coupling parameters $\sigma_1$ and $\sigma_2$ as shown in Figs. \ref{fig:synchro-error-ring-chimera} and \ref{fig:synchro-error-star-chimera}. It is evident that while the synchronization in RDC networks occurs for some higher values of $\sigma_1$ and $\sigma_2$, i.e.,  $26\lesssim\sigma_1\lesssim40$ [Fig. \ref{fig:synchro-error-ring-chimera} (a)] and $20\lesssim\sigma_2\lesssim35$ [Fig. \ref{fig:synchro-error-ring-chimera} (b)], the same  for SC networks occurs at lower values of $\sigma_1$ and $\sigma_2$, i.e., $2.3\lesssim\sigma_1\lesssim6.6$ [Fig. \ref{fig:synchro-error-star-chimera} (a)] and $0<\sigma_2\lesssim6$ [Fig. \ref{fig:synchro-error-star-chimera} (b)]. In the other intervals, the chimera states may be formed.  
%%%%%%%%%%%%%%%%%%%%%%%%%%%%%%%%
%%%%%%%%%%%%%%%%%%%%%%%%%%%%%%%%%%%%%%%%%%%%%%%
%%%%%%%%%%%%%%%%%%%%%%%%%%%%%%%%%%%%%%
\section{Discussion and Conclusion}
We have proposed a non-autonomous dynamical system for an  optically modulated  CO$_2$ laser  in presence of electro-optic feedback beams: one in the form of a small-amplitude time-dependent perturbation and  the other a negative feedback of subharmonic components of the laser intensity signal. A numerical study of the dynamical system reveals that the feedback beams can indeed drive the CO$_2$ lasers into hyperchaotic states. The periodic, multi-periodic and chaotic states of the laser are also found to coexist for different values of the key parameters $B$, $\epsilon$ and $R$ associated with, respectively, the bias voltage, the modulation depth of the feedback beam and the gain from the feedback loop.  A network of a finite number of hyperchaotic coupled CO$_2$ lasers as its nodes or oscillators is constructed with linear coupling strengths, and its synchronization    is studied by the method of master stability  function   \cite{pecora1998}.  It is found that a network of at most $24$ identical
oscillators may   be fully synchronized (where $\gamma_{\text{max}}<0$)   both in the star   and    ring of diffusively coupled  networks.   In both these cases, the    synchronization errors,  as obtained from numerical simulation of $24$ coupled hyperchaotic CO$_2$ lasers, are    $\lesssim10^{-6}$. It is also found that,  while the lower coupling strengths $\sim o(1)$ can lead to a synchronous state of star networks, relatively higher coupling strengths are required for the nearest-neighbor diffusive (ring) networks. 
\par
It is to be noted that in our network model, we have considered only the ring of diffusively coupled (RDC) and star coupling (SC) networks. However,  different other networks can be constructed, e.g., by combining both these ring and star networks, and   their synchronization can be studied following the present stability analysis. In this case, the matrix $G$, and so $G_1$ and $G_2$ may not be the same as the present ones.  Also, the numbers $n=20,~24$ and $28$ does not have any particular meaning. These   are  considered arbitrarily. In fact, one can consider any number of modes to examine the synchronization in networks.  In our theory, we have seen that the synchronization error becomes higher $(\gtrsim10^{-6})$ for $n>24$.  For illustrations, we have examined the synchronization with different number of nodes, i.e.,  $n=20,~24$ and $28$  to obtain the synchronization errors as $\sim 10^{-12},~10^{-6}$ and $10^{-5}$ respectively. Furthermore, the chimera states of the networks may coexist in  some intervals of time and the coupling parameters $\sigma_1$ and $\sigma_2$ where the networks are not synchronized. This   means that the synchronization occurs only in some specific ranges of values of $\sigma_1$ and $\sigma_2$, namely,   $26\lesssim\sigma_1\lesssim40$   and $20\lesssim\sigma_2\lesssim35$  for RDC networks, and    $2.3\lesssim\sigma_1\lesssim6.6$   and $0<\sigma_2\lesssim6$ for SC networks.   However, the detailed discussion about the formation of chimera states is beyond the scope of the present work.   
\par
In the present network model, each node is connected to independent but identical CO$_2$ lasers which exhibit hyperchaos. Since in the process of synchronization, all the connected nodes behave in a similar manner, one can  use this property of   RDC and SC networks  to produce a secure networking communication system in which the sender hides a message within the hyperchaotic signal that can only be recovered by the receiver at the synchronized state. Such an approach has been extensively applied in many secure communications, especially in optical chaos communication systems because of the added security and the speed of optical communications \cite{parlitz1996}.
In this way, one can also  develop the public key cryptography scheme using the process of chaos synchronization  which reduces the difficulties of the problems of key distribution
in an encryption process \cite{banerjee2011,roy2019,li2019b}.
%%%%%%%%%%%%%%
\par
Recently,  a different approach other than the MSF scheme has been proposed in Ref. \cite{abarbanel2008}, however, the method has some limitations, and we are not sure
whether this method can be successfully applied to the present model to examine  the synchronization in networks of CO$_2$ lasers. It requires further investigation which is beyond the scope of the present work.     Also, the theory of phase synchronization   has been extended to chaotic model oscillators   and to several laser experiments. In this context, the synchronization in presence of noise   which  usually has a destructive effect on phase synchronization by inducing phase slips and shrinking the synchronization region may be interesting to study \cite{zhou2003}. The  inclusion of the time delay effect of the light being fed back to the cavity, and also   the laser field phase in the present model may be other problems of interest. 
\par
To conclude, the synchronization phenomena in our network of hyperchaotic lasers may be applicable to neural networks which are information processing paradigms inspired by the way biological neural systems process data.  
 %%%%%%%%%%%%%%%%%%%%%%%%%%%%%%%%%%%%%%%
\section*{Acknowledgments}
We sincerely thank  Amitava Bandyopadhyay of Department of Physics, Visva-Bharati, India for  helpful suggestions in designing the diagram (Fig. 1) for CO$_2$ lasers. One of us  (A. P. M.) is supported by a Major Research Project sponsored by Science and Engineering Research Board (SERB), Government of India with sanction order no. CRG/2018/004475.

\end{document}